\documentclass[conference]{IEEEtran}

\begin{document}

\title{Information Retrieval Model: A Social Network Extraction Perspective}

\author{\IEEEauthorblockN{Mahyuddin K. M. Nasution}
\IEEEauthorblockA{Knowledge Technology Research Group,\\Faculty of Information Science \& Technology\\
Universiti Kebangsaan Malaysia\\
Bangi 43600 UKM Selangor Malaysia\\
Email: mahyunst@yahoo.com}
\and
\IEEEauthorblockN{Shahrul Azman Noah}
\IEEEauthorblockA{Knowledge Technology Research Group,\\Faculty of Information Science \& Technology\\
Universiti Kebangsaan Malaysia\\
Bangi 43600 UKM Selangor Malaysia\\
Email: samn@ftsm.ukm.my}}

\maketitle

\begin{abstract}
Future Information Retrieval, especially in connection with the internet, will incorporate the content descriptions that are generated with social network extraction technologies and preferably incorporate the probability theory for assigning the semantic. Although there is an increasing interest about social network extraction, but a little of them has a significant impact to infomation retrieval. Therefore this paper proposes a model of information retrieval from the social network extraction. 


\end{abstract}
 
\IEEEpeerreviewmaketitle

\section{Introduction}

Information Retrieval (IR) is concerned with answering information needs as accurately as possible \cite{cooper1971,manning2009}. The information is found in ever-growing documents collections such as web pages, for which information extraction algorithms are being developed on a large scale \cite{guodong2007}. Social network extraction establishes a technology to identify and describe special content: entities and their relations, that is represetation of web pages and query that are enriched with semantic information of their respective content \cite{mika2005,mika2006,matsuo2007,nasution2010}. However, there is little attention about the application of social network to IR \cite{cross2001}. Although, it will give rise to adapted and advanced retrieval model \cite{borgatti2003}. 

We assure that this technology will become an important part of IR, and IR as an application where the extracted social network contributes to a more refined representation of web pages and query. It is our goal in this paper, a model is defined by the query and web pages representations and by the function that estimates the relevance of a web page and a query based on relations among entities pairly.

\section{Related Work}
Well known models of IR are the Boolean, vector space, probabilistic, fuzzy and imaging. These have been studied in detail and implementation for experimentation, as well as, commercial purposes. Nevertheless, the known limitations of these models have caused researchers to propose new models. One such model is the logical model for IR \cite{bruza2000,crestani1998,dominich2000,lalmas1998}. In recent years there have been several attemps to define a logic for IR along the so-called logical approach, initiated by the pioneer work \cite{cooper1971} and given decisive impulse by two related works \cite{rijsbergen1986,rijsbergen1989}. Logical IR models were studied to provide a rich and uniform representation of information and its semantics, with the aim to improve retrieval effectiveness. 

In line with those logic studies, there are formal researches dealing with Imaging in IR \cite{crestani1995}. This idea explicitly proposed \cite{rijsbergen1989} and implemented \cite{amati1992,sembok1993} mainly to solve uncertainty problem in IR. They enable a more complex definitions of relevance than other IR models. Using similar approach, our model also based on logical-uncertainty and probability theory for enhancing IR by using the approaches of social network extraction, where the output of IR system based on score of relevance so that documents can be sorted according to relevance to the query, but future models will incorporate the content descriptions that are generated with IR and preferably incorporate the probabilistic nature of the assignments of the semantics.

\section{The Concept and Motivation}
Social networks extraction is the web pages (or documents) based process developed in the framework of modal relation \cite{nasution2010}. In the semantic web, there is one research stream of social network extraction depends heavily on the co-occurrence as modal relation by utilizing the Cartesian product for clustering on the space of events \cite{mika2005,matsuo2007}.\\

\noindent
{\it Definition 3.1} \cite{nasution2011} Let $V\ne\emptyset$ is a set of nodes and $E$ is a set of edges. The social network extraction with the exertions $\xi$ and $\zeta$ for acquiring rich and trusted social network is
\[
SN = \langle V,E,A,R,Z,\xi,\zeta\rangle
\]
that satisfies the following conditions
\begin{enumerate}
\item $\xi(1:1) : A \rightarrow V$, $v=\xi(a)$, $\forall a\in A$ $\exists !v\in V$, where $A$ is a set of actors.
\item $\zeta : R\rightarrow E$ so that $e_j=\zeta(r_k(a,b)) = \zeta(Z_a\cap Z_b)$, $e_j\in E$, $r_k\in R$, $\forall a,b\in A$, $Z_a,Z_b,Z_a\cap Z_b\subseteq Z$, where $Z$ is a set of attributes.
\end{enumerate}

In Definition 3.1, the use of social network data in IR is a motivation for computing the importance of an actor in the social network that is extracted from the web documents and for using this relevance to compute the importance a web page. It also enables the relations among the social actors (entities) allow us to create and maintain an aggregate of close web-documents.

Information Retrieval is a knowledge technology concerned with the effective and efficient retrieval of information for the subsequent use by interested parties. Information retrieval typically involves the querying of unstructured or semi-structured information, the former referring to the content  of unstructured text (written or spoken), images, video and audio, the latter referring to well-defined metadata that are attached to the web pages especially. Like other technologies, IR can also be formulated using the logic \cite{chiaramella1992}. Logical reasoning is the essence of the defining paradigms and for understanding the phenomena in the space of events. In classical logic where inference is often associated with logical implication: a web page $\omega$ is relevant to a query $q$ if it implies the query, or in other words, if the query can be inferred from the web page, that is if $\omega\Rightarrow q$ (read "if $\omega$ then $q$") is true. A well-kown paradigm of querying a web page is by inputting key terms and matching them against the terms by which the web pages are indexed. The term is the words of the texts in case of a full text search we define as follows.\\ 

\noindent
{\it Definition 3.2} A term $t_x$ consists of at least one or a set of words in a pattern, or $t_x = (w_1,\dots,w_l)$, $l\leq k$, $k$ is a number of parameters representing word $w$, $l$ is number of tokens (vocabularies) in $t_x$, $|t_x| = k$ is size of $t_x$.\\

We use the term for defining the singleton event as follows.\\

\noindent
{\it Definition 3.3}
Let a set of Web pages indexed by search engine be $\Omega$. For each search term $t_a$, where $t_a\in \Sigma$, i.e., a set of singleton search term of search engine. A vector space $\Omega_a\subseteq\Omega$ is a singleton search engine event of web pages (\emph{singleton event}) that contain an occurrence (event) of $t_a\in\omega$. The probability of a singleton event $\Omega_a$ is 
\begin{equation}
P(t_a) = |\Omega_a|/|\Omega| \in [0,1] 
\end{equation}
where $|\Omega|$ is the cardinality of $\Omega$, and $|\Omega_a|\leq|\Omega|$.\\

Let $t_a$ represents an actor/entity name and $t_a$ is in the query, then $\omega\Rightarrow t_a$. The implication of $\omega\Rightarrow t_a$ is as an interpretation of exertion $\xi$ in Definition 3.1. However, logic by itself cannot fully model IR. In determining the relevance of a web page to a query, the success or failure of an implication relating the true or false values is not enough. Although, a web page consists of a set of statements, or $\{s_i|i=1,\dots,m\}$, but a collection of web pages cannot be considered as a consistent set of statements containing $t_a$. In fact, the web pages in the collection could and often do contradict each other in any particular logic, and not all the necessary knowledge is available. In case of name disambiguation, $\Omega_t = \{\omega_i|i=1,\dots,I\}$ is a set of web pages containing the names where a name could and often consist of different patterns of name tokens (first/middle/last names or in abbreviation). Together with growing the web on Internet, the presences of semantic relation such as synonymy and polysemy gave (a) different entities can share the same name, and (b) a single entity can be designated by multiple names. Therefore, the relationships between web pages and the entities in logic is uncertain, and degree of uncertainty measured by $P(\omega\Rightarrow t_a)$ and estimated by the conditional probability $P(t_a|\omega)$, a conditional events $(t_a\cap\omega)/\omega$. 

For singleton event the $\Omega$ be the space of events where $\omega\Rightarrow t_a$ is true, or
\[
\Omega_a(t_a) = \cases{1 & if $t_a$ is true at $\omega_a \in \Omega$\cr
0 & otherwise,\cr}
\]
but $\Omega_a(t_a) = 1$ also for $w_1,w_2,\dots,w_l \in t_a$ without their pattern (such as $t_a = (w_1,w_2,\dots,w_l)$) is true at any $\omega\in\Omega$. Each of the returned web pages may contain many relevant information and even some irrelevant ones. Thus, 
\begin{equation}
|\Omega_a| = \sum_\Omega (\Omega_a(t_a)=1) \geq \sum_\Omega (\omega\Rightarrow t_a).
\end{equation}
$\sum_\Omega(\omega\Rightarrow t_a)$ is the number of web pages containing $t_a$. Therefore, to assemble the relevant information mentioned in web pages, we can identify the most relevant information amongst those mentioned in the top $n$ results based on a insight: how often the information is mentioned across the top results also provides important hint about its relevance to the query. The "relevant information" means that if a user of an IR system has an information need, such as relevance is defined as \emph{logical consequence} \cite{cooper1971}, whereby the query is represented by statement and its negation that consist of premiss set and minimal premiss set, i.e., a statement as a logical consequence of subset of sentences, and a statement is one that is as small as possible in the sense that if any of its members were deleted. It enables we can define a boundary $\beta_a$ such that 
\begin{equation}
\sum_\Omega (\omega\Rightarrow t_a) \simeq |\Omega_a|-\beta_a.
\end{equation}
or based on Definition 3.1 and Definition 3.2, we have
\begin{equation}
\begin{array}{rcl}
v &=& \xi(a)\cr
  &=& P(\sum_\Omega (\omega\Rightarrow t_a))\cr
  &\simeq& (|\Omega_a|-\beta_a)/|\Omega|\cr
\end{array}
\end{equation}
It also proves that the following proposition.\\

\noindent
{\it Proposition 3.1}
A web page is relevant to an entity information need, $\omega\equiv t_a$, if web page contains at least one sentence where there is $t_a$ in name disambiguation condition ($\exists !s, t_a\in s$).\\

For extracting the social networks from web pages, to accompany singleton we define a doubleton event.\\

\noindent
{\it Definition 3.4}
Let a set of Web pages indexed by search engine be $\Omega$. We assume $t_a$ and $t_b$ are search terms, $t_a\ne t_b$, $t_a,t_b\in\Sigma$, where $\Sigma$ is a set of singleton search term of search engine. A doubleton search term is $\{\{t_a,t_b\}:t_a,t_b\in\Sigma\}$ and its vector space denoted by $\Omega_a\cap\Omega_b\subseteq\Omega$ is a doubleton search engine event of web pages (\emph{doubleton event}) that contain a co-occurrence of $t_a$ and $t_b$ such that $t_a,t_b\in\omega_a$ and $t_a,t_b\in\omega_b$. Probability of a doubleton event $\Omega_a\cap\Omega_b$ is
\begin{equation}
P(t_a,t_b) = |\Omega_a\cap\Omega_b|/|\Omega|\in[0,1]
\end{equation}
where $\Omega_a,\Omega_b,\Omega_a\cap\Omega_b\subseteq\Omega$.\\
 
In social network, the researchers leverage the top relevant web pages and relationship between web pages to identify the most relevant entities \cite{matsuo2007}, or generate a relationship between one entity to another based on relevant web pages. It means that also how close one web page to another.\\

\noindent
{\it Theorem 3.1}
The web pages are relevant to an information need for the social network if it contains at least one sentence which is relevant to the relation between two entities.\\

We borrow the use of logic in imaging \cite{crestani1995} to prove this theorem. Let two terms $t_a\ne t_b$ for different entities, $\omega\Rightarrow t_a\wedge t_b$. Let $\Omega_a$ and $\Omega_b$ are the spaces of event for $\omega_a \Rightarrow t_a$ and $\omega_b \Rightarrow t_b$ are true, respectively. $\Omega_a$ be most similar to $\Omega_b$ where $t_a$ is true, then $t_a\Rightarrow t_b$ will be true at $\Omega_b$ if and only if $t_b$ is true at $\Omega_a$, that is $\Omega_b(t_a) = 1$ if $t_a$ is true at $\Omega_b$ then we have 
\begin{equation}
\Omega_b(t_a\Rightarrow t_b) = \Omega_a(t_b)
\end{equation}
where $\Omega_a(t_b) = 1$ if $t_b$ is true at $\Omega_a$. Similarly, based on symmetry of similarity, we obtain
\begin{equation}
\Omega_a(t_b\Rightarrow t_a) = \Omega_b(t_a)
\end{equation}
If $\omega_a \in \Omega_a$, $\omega_a\in\Omega_b$ and $\omega_b\in\Omega_b$, $\omega_b\in\Omega_a$, then it applies that $t_a$ and $t_b$ be co-occurrence in $\Omega$, or $\Omega_a\cap\Omega_b(t_a\Rightarrow t_b) = 1$, where $\Omega_a,\Omega_b\subseteq\Omega$, and
\[
\begin{array}{rcl}
\Omega_a\cap\Omega_b(t_a\Rightarrow t_b) &=& \Omega_a(t_a\Rightarrow t_b)\wedge\Omega_b(t_b\Rightarrow t_a)\cr
 &=& 1 \wedge 1\cr
 &=& 1,
\end{array}
\]
$\Omega_a\cap\Omega_b$ is the space of event if $\omega_a \Rightarrow t_a\wedge t_b$ and $\omega_b \Rightarrow t_a\wedge t_b$ are true.

Similar to singleton event, based on Theorem 3.1 the doubleton event be space of event where $\omega\Rightarrow t_a\wedge t_b$ is true. Thus,
\begin{equation}
\begin{array}{rcl}
|\Omega_{ab}| &=& |\Omega_a\cap\Omega_b|\cr
 &=& \sum_\Omega(\Omega_{ab}(t_a,t_b)=1) \geq \sum_\Omega(\omega\Rightarrow t_a\wedge t_b).\cr
\end{array}
\end{equation}
or with a boundary $\beta_{ab}$,
\begin{equation}
\sum_\Omega(\omega\Rightarrow t_a\wedge t_b) \simeq |\Omega_a\cap\Omega_b|-\beta_{ab},
\end{equation}
or from Definiton 3.1 and Definition 3.4, we obtain
\begin{equation}
\begin{array}{rcl}
e_{ab} &=& \zeta(a,b)\cr
  &=& P(\sum_\Omega(\omega\Rightarrow t_a\wedge t_b))\cr
\end{array}
\end{equation}
or in Jaccard coefficient
\begin{equation}
e_{ab} = \frac{|\Omega_a\cap\Omega_b|-\beta_{ab}}{|\Omega_a| + |\Omega_b| - |\Omega_a\cap\Omega_b|
- \beta_a - \beta_b + \beta_{ab}} 
\end{equation}
Because of $|\Omega_a\cap\Omega_b| \leq \Omega_a$ or $|\Omega_a\cap\Omega_b| \leq \Omega_b$, $\beta_{ab}$ is less than or equal to $\beta_a$ or $\beta_b$.

The doubleton event is an attempt in order to two entities are related when they are often mentioned in the same context, mainly author-coauthor relationship at their academic papers. This attempt is to be distinguished from looser ones like, for instance, the vector space in which web pages are ranked according to a measure of similarity with the query. Of course, the social network extraction based on a treatment of similarity, mainly to find out the strength relations operationally in a number of ways, and the similarity based models of IR generally lack the theoretical soundness of probabilities models. However, IR is able to search efficiently through huge amounts of data because it builds indexes from the web pages (documents), where all kinds of information needs that are very difficult to determine a priori. Moreover, The queries with the words do not always occur in relevant web pages. Thus, the query expansion with synonym and related terms (or have relation) is one popular alternative, primarily enhancing the recall of the results of the search, but Eqs. (10) and (11) mean that the relation is uncertain also. Therefore, although a measure of similarity cannot be directly interpretable as a probability, we can use probabilistic inference to cope with uncertainty a relation.

\section{Model of IR}

Social network reflects a shift from the individualism common towards a social structure, or an exchange from information individually to the information of relations, where the fundamental units were defined, i.e., the relations between entities. Relations are characterized by content, direction and strength. The content of a relation refers to the resource that is exchanged, a relation can be directed or undirected, and relation also differ in strength \cite{nasution2011b}. 

Suppose we have a set of possible relations $R$. Definition 3.3 means that $t_a\Rightarrow \omega$, $\forall\omega\in\Omega_a$ such that $\omega$ connect to another in singleton event by $t_a$. In this case, the space of event $\Omega_a$ be in a space of relation $\rho$ (or there are a web pages network) where $\xi$ generates $\rho$, and degree of uncertainty can be measured by Eq. (4). For all $\rho_x\in R$ be the space of relations tie to $\rho_y\in R$ where the space of event $\Omega_a$ is true, and $\Omega_a\Rightarrow\Omega_b$ will be true at $\rho_y$ if and only if $\Omega_b$ is true at $\rho_x$. Or
\[
\rho_y(\Omega_a)=\cases{1 & if $\Omega_a$ is true at $\rho_y$\cr
0 & otherwise\cr}
\]
and
\[
\rho_x(\Omega_b)=\cases{1 & if $\Omega_b$ is true at $\rho_x$\cr
0 & otherwise\cr}
\]
then we obtain
\begin{equation}
\rho_y(\Omega_a\Rightarrow\Omega_b) = \rho_x(\Omega_b).
\end{equation}
Similarly, if $\rho_x$ and $\rho_y$ are the spaces of symmetry relations, we have
\begin{equation}
\rho_x(\Omega_b\Rightarrow\Omega_a) = \rho_y(\Omega_a).
\end{equation}
From Eqs. (6) and (7), we obtain
\[
\begin{array}{rcl}
\rho_x(\Omega_b(t_a\Rightarrow t_b)\Rightarrow\Omega_a(t_b\Rightarrow t_a)) &=& \rho_y(\Omega_a(t_b)\Rightarrow\Omega_b(t_a))\cr
&=& \rho_y(\Omega_a\Rightarrow\Omega_b)\cr
&=& \rho_x(\Omega_b)\cr
\end{array}
\]
It concludes into the following statement.\\

\noindent
{\it Theorem 3.2}
If a social network can be extracted from web pages (or collection of documents), then web pages as space of events in the space in relations is a document network that represent the document collection.\\

In IR model, this theorem is to provide that $\omega\Rightarrow t_a\wedge t_b$ is also $t_a\wedge t_b\Rightarrow\omega$ in space of relation $\rho \in R$, where $\xi$ and $\zeta$ (social network extraction) generate $\rho$, i.e. first is in web pages networks and second is in actors networks.

A treatment of probability of either event or relation spaces based on a probability distribution over the set of possible events $\Omega$ or set of possible relations $R$, respectively. There are $\sum_\Omega P(\omega) = 1$ and $\sum_R P(\Omega) = 1$.
By imaging we have 
\begin{equation}
P(\omega\Rightarrow q) = P_\omega(q) = \sum_\Omega P(\omega)\Omega_\omega(q),
\end{equation}
where $\Omega_\omega(q) = 1$ if $q$ is true at $\Omega_\omega$, $\Omega_\omega(q) = 0$ otherwise.
\begin{equation}
P(\rho\Rightarrow\omega) = P_\rho(\omega) = \sum_R P(\rho)R_\rho(\omega),
\end{equation}
if $\omega$ is true at $R_\rho$ then $R_\rho(\omega) = 1$ else $R_\rho(\omega) = 0$. Probability of relation in a social network, based on uncertainty of Eqs. (3) and (8) such as Eq. (11), depends on satisfying a threshold that derived by the boundaries. In an inference network, the truth value of a node depends only upon the truth values of its parents. To evaluate the strength of an inference chain going from one web page to the query we set the web page node $\omega_i$ to true and evaluate $P(q_k=true|\omega_i=true)$. This gives us an estimate of $P(\omega_i\Rightarrow q_k)$, where $P(\omega\Rightarrow\rho)$ is document network represents the collection of web pages such as singleton events $\Omega_a,\Omega_b\subseteq\Omega$ or doubleton event $\Omega_a\cap\Omega_b\subseteq\Omega$. While a query network is built for each information need and can be modified and extended during each session by the user in a interaction and dynamic way, in an inference network we have
\begin{equation}
P(\rho\Rightarrow q) = P_\rho(q) = \sum_R P(\rho)R_\rho(q),
\end{equation}
where
\[
R_\rho(q) =\cases{1 & if $q$ is true at $R_\rho$,\cr 0 & otherwise.\cr}
\]
We substitute Eqs. (15) and (16) into Eq. (14) as follows
\[
\begin{array}{rcl}
P(\omega\Rightarrow t_a\wedge t_b) &=& P(\omega\Rightarrow q)\cr 
 &=& P((\rho\Rightarrow\omega)\Rightarrow(\rho\Rightarrow q))\cr&=& P(\rho\Rightarrow\omega)\Rightarrow P(\rho\Rightarrow q)\cr
&=& P_\rho(\omega) \Rightarrow P_\rho(q)\cr
&=& \sum_R P(\rho)R_\rho(\omega)\Rightarrow \sum_R P(\rho)R_\rho(q)\cr
&=& \sum_R P(\rho)R_\rho(q) \Rightarrow \sum_R P(\rho)R_\rho(\omega)\cr
&=& P_\rho(q)\Rightarrow P_\rho(\omega)\cr
&=& P(\rho\Rightarrow q)\Rightarrow P(\rho\Rightarrow\omega)\cr
&=& P((\rho\Rightarrow q)\Rightarrow(\rho\Rightarrow\omega))\cr
&=& P(q\Rightarrow\omega)\cr
&=& P(t_a\wedge t_b\Rightarrow \omega)\cr
\end{array}
\]
If $t_a,t_b$ in $q$, then a web page $\omega$ relevant to a relation. Therefore, the evidence is two or more terms $t_j$ or relations $\rho_j$ together are relevant for the web pages. Difference combinations of terms in query can be activated and their relevance can be computed on social networks of their recorded entities.

\section{Conclusion and Future Work}
The social network will usually be manageable in terms (represent entities) of computational complexity. It is also possible to activate not one candidate entity or passage when computing relevance, but considering a number of combinations of relations between entities to be active and to compute the relevance of the set. One can compute the event that two or more entities $t_i, i=1,\dots, n$ or relations $\rho_j, j=1,\dots,m$ together are relevant for the query. The terms of the entities can be linked to different concepts, which, for instance represent corefering entities or events. The relations can be extracted from different web pages, and all possible of web pages as an answer to the question is computational feasible or not. Our near future work is to further experiment the proposed method and look into the possibility of enhancing IR performance by using social networks.

\end{document}